\documentclass[%
   draft=false,      % fertiges Dokument
   fontsize=11pt,
   normalheadings,   % normale Überschriften
   english,           % wird an andere Pakete weitergereicht
   paper=a4,
   BCOR5mm,          % Zusaetzlicher Rand auf der Innenseite
   DIV11,            % Seitengroesse (siehe Koma Skript Dokumentation !)
   1.1headlines,     % Zeilenanzahl der Kopfzeilen
   headinclude=false,% Kopf nicht einbeziehen
   footinclude=false,% Fuss nihct einbeziehen
   mpinclude=false,  % Margin nicht einbeziehen
   pagesize,         % Schreibt die Papiergroesse in die Datei.
   twoside,          % Seitenraender für zweiseitiges Layout
   openright,        % Kapitel beginnen immer auf der rechten Seite
   cleardoublepage=empty,    % leere, linke Seite mit Seitenstil 'empty'
   titlepage,        % Titel als einzelne Seite ('titlepage' Umgebung)
   parskip=half+,
   headsepline,      % Linie unter Kolumnentitel
   nochapterprefix,  % keine Ausgabe von 'Kapitel:'
   liststotoc,      % Tabellen & Abbildungsverzeichnis ins TOC
   bibtotoc,         % Bibliographie ins TOC
   tocindent,        % eingereuckte Gliederung
   listsindent,      % eingereuckte LOT, LOF
   pointlessnumbers, % Überschriftnummerierung ohne Punkt, siehe DUDEN !
   fleqn,            % Formeln werden linksbuendig angezeigt
]{scrbook}%     Klassen: scrartcl, scrreprt, scrbook

\usepackage{scrpage2}
\ohead[]{\pagemark}
\ihead[]{\headmark}
\ofoot[\pagemark]{}

\usepackage[english]{babel}

\usepackage[T1]{fontenc}
\usepackage{fourier}
\usepackage{courier}
\usepackage{color}

\definecolor{gray2}{gray}{0.2}
\definecolor{gray4}{gray}{0.4}
\definecolor{gray6}{gray}{0.6}
\definecolor{gray8}{gray}{0.8}
\definecolor{gray9}{gray}{0.9}
\definecolor{cLink}{rgb}{0.0, 0.5, 0.0}
\definecolor{cCite}{rgb}{0.0, 0.5, 0.5}
\definecolor{cFile}{rgb}{0.0, 0.0, 0.5}
\definecolor{cUrls}{rgb}{0.0, 0.0, 0.5}
\definecolor{accent1}{rgb}{0.0, 0.0, 0.7}
\definecolor{accent2}{rgb}{0.0, 0.7, 0.0}
\definecolor{accent3}{rgb}{0.7, 0.0, 0.0}

\usepackage{nameref}
\usepackage[bookmarks=false]{hyperref}
\let\Oldnameref\nameref
\renewcommand{\nameref}[1]{\textit{\Oldnameref{#1}}}

\usepackage[babel,english=american]{csquotes}		% american
\usepackage[intoc]{nomencl}
\renewcommand{\nomname}{Settings}
\renewcommand{\nomlabel}[1]{\textbf{#1} \dotfill}
\makenomenclature

\makeatletter
  {%
  \endlist
  \nompostamble}
\makeatother

\usepackage[
   bottom,      % Footnotes appear always on bottom. This is necessary especially when floats are used
   stable,      % Make footnotes stable in section titles
   perpage,     % Reset on each page
   multiple,    % rearrange multiple footnotes intelligent in the text.
]{footmisc}

\usepackage{listings}
\usepackage{graphicx}       % graphicx with PDFTeX driver
\graphicspath{{images}{graphics/}}

\usepackage[table]{xcolor}
\newcommand{\bParams}{\textbf{Parameters}~\\ \rowcolors{1}{gray9}{gray8} \begin{tabular*}{\textwidth}{p{.3\textwidth}p{.6\textwidth}}}
\newcommand{\eParams}{\end{tabular*}}

\let\Oldsec\section
\renewcommand{\section}[1]{\Oldsec{#1}\label{#1}}

\newcommand{\bPrv}[1]{\stepcounter{subsection} \subsection*{\thesubsection ~\textcolor{accent3}{\texttt{PRIVATE}}  #1}}

\newcommand{\jString}{\textsc{String}~}
\newcommand{\jStringa}{\textsc{String[]~}}

\newcommand{\jBytea}{\textsc{byte[]~}}
\newcommand{\jInt}{\textsc{int~}}
\newcommand{\jInta}{\textsc{int[]~}}
\newcommand{\jObject}{\textsc{Object~}}

\newcommand{\J}[1]{\textsc{#1~}}

\newcommand{\constructor}[4]{
\subsection{\textcolor{accent1}{\texttt{CONSTRUCTOR}}  #1}
	\textbf{Description}\\
	#2
	\\\\
	\textbf{Settings}\\
	#3
	 
	\bParams
		#4
	\eParams
}

\newcommand{\publicMethod}[5]{
\subsection{\textcolor{accent2}{\texttt{PUBLIC}} #1}
	\textbf{Description}\\
	#2
	\\\\
	\textbf{Settings}\\
	#3
	 
	\bParams
		#4
	\eParams
	\\\\
	\textbf{Output}\\
	#5
}

\newcommand{\publicMethodNoOut}[4]{
\subsection{\textcolor{accent2}{\texttt{PUBLIC}} #1}
	\textbf{Description}\\
	#2
	\\\\
	\textbf{Settings}\\
	#3
	 
	\bParams
		#4
	\eParams
}

\newcommand{\privateMethod}[5]{
\bPrv{#1}
	\textbf{Description}\\
	#2
	\\\\
	\textbf{Settings}\\
	#3
	
	\bParams
		#4
	\eParams
	\\\\
	\textbf{Output}\\
	#5
}

\newcommand{\privateMethodNoOut}[4]{
\bPrv{#1}
	\textbf{Description}\\
	#2
	\\\\
	\textbf{Settings}\\
	#3
	
	\bParams
		#4
	\eParams
}

\newcommand{\ethemba}{
\htmladdnormallink{\textcolor{cLink}{ethemba}}{http://www.ethemba.info/}
}

\newcommand{\ethembaG}{
\htmladdnormallink{\includegraphics[width=\textwidth]{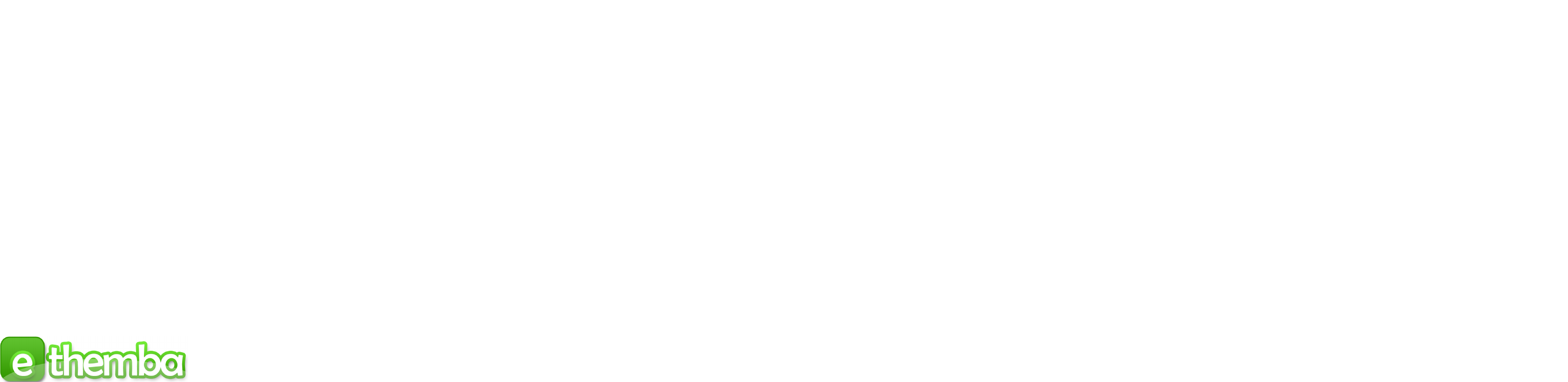}}{http://www.ethemba.info/}
}

\begin{document}

	%
% Title-Page
%

\begin{titlepage}
	\thispagestyle{empty}
			
	~
	\vspace{10em}
	\center
	\htmladdnormallink{\includegraphics[width=1.00\textwidth]{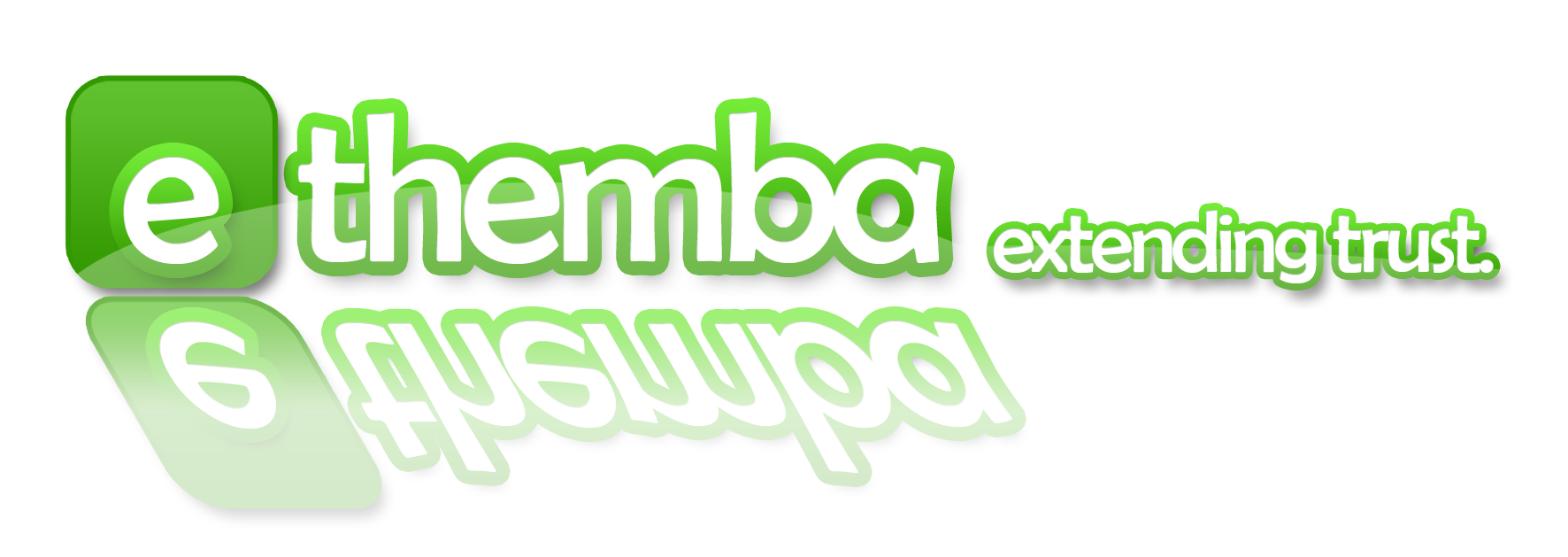}}{http://www.ethemba.info/}
	~\\
	\vspace{6em}
	\Large{\textbf{E}themba \textbf{T}rusted \textbf{H}ost \textbf{E}nvironment \textbf{M}ainly \textbf{B}ased on \textbf{A}ttestation}\\
	\vspace{0.3em}
	\large{\textit{a framework and demonstrator for TPM applications}}\\
	\vspace{5em}
	\Large{Andreas Brett}\\
		\normalsize{\texttt{\href{mailto:andreas\_brett@ethemba.info}{andreas\_brett@ethemba.info} } }\\~\\
	\Large{Andreas Leicher}\\
	  \normalsize{\texttt{ \href{mailto:andreas\_leicher@ethemba.info}{andreas\_leicher@ethemba.info}} }\\
	\vspace{3em}
	%\large{\today}
	\large{December 24, 2008}
\end{titlepage}

	\thispagestyle{empty}

\begin{flushright}
	\textit{This work was supported by}\\
	\textbf{Dr. Andreas U. Schmidt} (CREATE-NET)\\
	\textbf{Nicolai Kuntze} (Fraunhofer-SIT)
\end{flushright}

	\cleardoublepage
	\pdfbookmark[1]{Contents}{toc} 
	\tableofcontents

	\chapter{apps}
		\textsc{de.fraunhofer.sit.tc.ethemba.apps}\\\\
		This package contains applications for TPM maintenance on the one hand and client- and server-applications
		that implement \textbf{AIK-Certification} and \textbf{Remote-Attestation} on the other hand.
		
		\begin{itemize}
			\item \nameref{TakeOwnership} and \nameref{ClearOwnership} permit TPM activation capabilities
			\item \nameref{ManageKnownHashesList} helps managing hash lists, that define trusted applications used
			in \textbf{Remote-Attestation} on the server-side
			\item \nameref{PCAserver} and \nameref{PCAclient} implement \textbf{AIK-Certification}
			\item \nameref{RAserver} and \nameref{RAclient} implement \textbf{Remote-Attestation}
		\end{itemize}
		\section{ClearOwnership}
\subsection{Description}
	ClearOwnership clears the ownership of the TPM. The code is mainly based on the original jTpmtools (see \nameref{trustedjava}) application.
	It can be used in the process of resetting the TPM to its initial state. A reboot might be required for changes to take effect.	The
	current owner password is supplied as command-line parameter.
	
	For convenience and tests, the switch \textbf{/f} provides a fixed mode, reading the globally set owner password.
	
	\textbf{Note:} In our test scenario, routing the TPM-Emulator into a virtualized sub-system, we were able to clear the ownership \textbf{without} supplying the
	correct owner password. It is still to be clarified, if this is an issue with the emulated environment or the implementation in jTSS.

%
% main
%
\publicMethod
%title
{main}
%description
{Called when used in command-line.}
%settings
{OwnerPwd}
%parameters
{\jStringa args & [1] owner password or globally set owner password when used in fixed mode ('/f').}
%output
{If no command line parameters are given, a usage information is displayed.}
		\section{TakeOwnership}
\subsection{Description}
	TakeOwnership allows taking ownership of the TPM. The owner password is set, a new SRK is generated inside the TPM and the
	SRK password is set. The code is mainly based on the original jTpmtools (see \nameref{trustedjava}) application.
	
	The owner and SRK passwords are supplied as command-line parameters. For convenience and tests, the switch '/f' provides a
	fixed mode, reading the globally set owner and SRK passwords.
 	
%
% main
%
\publicMethod
%title
{main}
%description
{Called when used in command-line.}
%settings
{OwnerPwd, SRKPwd}
%parameters
{\jStringa args & [1] owner password or '/f' for fixed mode 

[2] srk password}
%output
{If no command line parameters are given, a usage information is displayed.}
		\include{CreateServerKeyPair}
		\section{ManageKnownHashesList}

\subsection{Description}
	ManageKnownHashesList takes an IMA-formatted file as input and converts it to a \nameref{KHL}. This
	application can be used to create and maintain a database of known hashes. Used in append-mode (command
	line parameter /a), the contents of the input will be appended to the database. Used in overwrite-mode
	(command line parameter /o), the contents of the input will overwrite an existing database. If no command
	line parameter is given, a management console is presented to the user, allowing for \textit{view}, \textit{search}
	and \textit{remove} entries of the database.

%
% main
%
\publicMethod
%title
{main}
%description
{Called when used in command-line.}
%settings
{RAServer\_KnownHashesList}
%parameters
{\jStringa args & [1] /a or /o to enable append- or overwrite-mode (optional)
 
 [2] filename of IMA-Measurement-File to be used when [1] is set}
%output
{If no command line parameters are given, a management console providing \textit{view}, \textit{search} and 
\textit{remove} functions is presented to the user. Otherwise the new or updated database of known hashes is stored in
RAServer\_KnownHashesList.}

%
% private append
%
\privateMethodNoOut
%title
{append}
%description
{appends contents of given IMA-Measurement-File to existing database.}
%settings
{RAServer\_KnownHashesList}
%parameters
{\textsc{String} file & IMA-Measurement-File}

%
% private overwrite
%
\privateMethodNoOut
%title
{overwrite}
%description
{overwrites existing database with contents of given IMA-Measurement-File.}
%settings
{RAServer\_KnownHashesList}
%parameters
{\textsc{String} file & IMA-Measurement-File}

%
% private view
%
\privateMethodNoOut
%title
{view}
%description
{displays the entries contained in the KnownHashesList pagewise (blocks of 10 entries)}
%settings
{none}
%parameters
{none}

%
% private search
%
\privateMethodNoOut
%title
{search}
%description
{displays the entries contained in the KnownHashesList matching the given search string}
%settings
{none}
%parameters
{\jString s & search string to be used}

%
% private remove
%
\privateMethodNoOut
%title
{remove}
%description
{removes entries contained in the KnownHashesList matching the given search string}
%settings
{none}
%parameters
{\jString s & search string to be used}
		\section{PCAclient}
The \nameref{PCAclient} can be used to create and certify an AIK. The client creates a new AIK using the \textbf{TPM\_CollateIdentityRequest}. The public part of it is sent to a \nameref{PCAserver}, including the Endorsement Certificate. They are encrypted using the \nameref{PCAserver}'s public key. Upon receipt, the \nameref{PCAserver} can decrypt the contents and verify the Endorsement Certificate. The \nameref{PCAserver} then creates a random nonce and wraps it for the client.
The response contains two parts:
\begin{enumerate}
	\item a symmetric session key and the hash of the AIK public key, both encrypted with the EK public key
	\item the nonce, encrypted with the session key
\end{enumerate}
for details see \nameref{PrivacyCa} state3\_sub.

Using the \textbf{TPM\_ActivateIdentity} command, the TPM can decrypt the answer and the nonce is revealed to the client. Only the client with the TPM used for creation of the AIK is able to decrypt the nonce.
The client then sends the decrypted nonce back to the \nameref{PCAserver}.

Upon receipt and verification, the \nameref{PCAserver}~ will create a certificate for the AIK. It is then encrypted using an \nameref{AES} key. The \nameref{AES} key is wrapped for the client using the same scheme as for the nonce, thus the response consists of three parts:
\begin{enumerate}
 \item a symmetric session key and the hash of the AIK public key, both encrypted with the EK public key
 \item the \nameref{AES} key encrypted with the session key
 \item the AIK-Certificate, encrypted with the \nameref{AES} key
\end{enumerate}
for details see \nameref{PrivacyCa} state5\_sub.

The AIK is activated by the client via a \textbf{TPM\_ActivateIdentity} call. The \nameref{AES} key is decrypted and can be used to decrypt the AIK-Certificate.

The client then assigns a UUID to the AIK and stores it in the System Persistent Storage of jTSS. For later access, the key is registered in the \nameref{client.TpmKeyDB} using the provided AIK-label.
As final step the AIK-Certificate is stored in the \nameref{client.CertDB}.

\begin{figure}[htbp]
	\centering
		\includegraphics[width=1.00\textwidth]{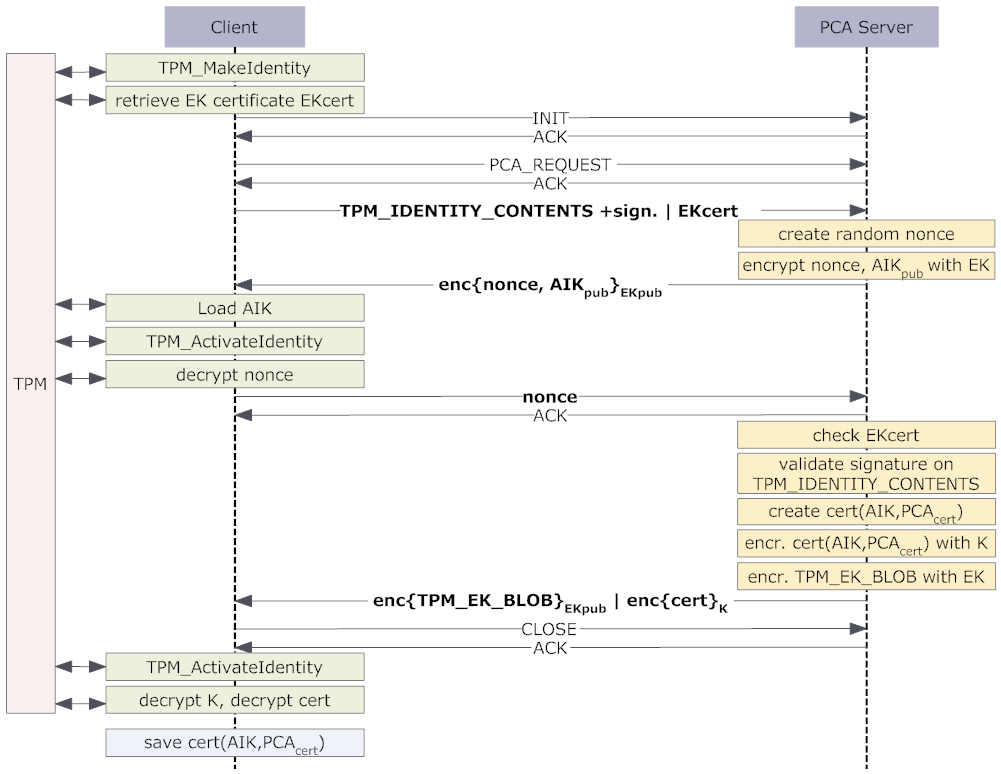}
	\caption{protocol diagram for AIK-Certification}
	\label{fig:small_PCA_protocol_Handshake}
\end{figure}

The client is separated into different methods representing each state of the protocol. The methods are called subsequently.

\bParams
owner password & needed to load the EK certificate / EK public key\\
SRK password & needed to store the new AIK\\
AIK password & will be needed when the AIK is accessed later on\\
AIK label & a label providing identification of the AIK. This label can later be used to load the AIK from \nameref{client.TpmKeyDB} and will also be included in the certificate.\\
PCA server IP address & IP address of the PCA server\\
PCA server port & port of the PCA server, see \nameref{Settings} for default value\\
\eParams

All required parameters can be set from the command-line, a fixed mode is provided via the switch '/f'. The required parameters will then be read from Settings.
		\section{PCAserver}
The server is separated into different methods representing each state of the AIK-Certification protocol. The methods are called subsequently.

$\rightarrow$ see \nameref{PCAclient} for a detailed description of the AIK-Certification protocol
		\section{RAclient}
The \nameref{RAclient} can be used to perform a \textbf{Remote-Attestation}. \textbf{Integrated Measurement Architecture} (IMA)
is used to provide runtime analysis of the client's system state. PCR-10 is constantly being extended with hashes of newly run
programs. A client is being attested by a Remote-Attestation-Server by sending an AIK-signed quote of PCR-10 altogether with the
corresponding AIK-Certificate and a \nameref{MeasurementList} to the server.

The quote is protected from replay-attacks by including a nonce, the Remote-Attestation-Server challenged the client with. By letting the
TPM quote PCR-10 internally, integrity of PCR-10 is ensured.

Upon receipt, the Remote-Attestation-Server performs several checks to attests the client's system state. First of all the AIK-Certificate
is checked for validity (timestamp, signature and trusted issuer). Then the submitted \nameref{MeasurementList} is validated by checking
each run application for known and trusted hashes. This results in a re-calculated PCR-10. At last the submitted quote is being validated
by checking its signature, the included nonce (for anti-replay!) and equality of the included PCR-10 and the re-calculated one.

This equality check is highly required to ensure an attacker did not trick the Remote-Attestation-Server by submitting a fake \nameref{MeasurementList}.

After verification the client will receive a certificate from \nameref{RAserver} certifying platform integrity. Note that this certificate has to be invalid after a short period of time, as the client's state might change very quickly.

The \nameref{RAclient} stores the Attestation-Certificate in the \nameref{client.CertDB} using a new UUID. The UUID for the Attestation-Certificate is based on the UUID of the AIK used in the protocol. It can be accessed for later use via the UUID \texttt{00000009-0008-0007-0605-}\textit{aikUuid.getNode()}.

\begin{figure}[htbp]
	\centering
		\includegraphics[width=1.00\textwidth]{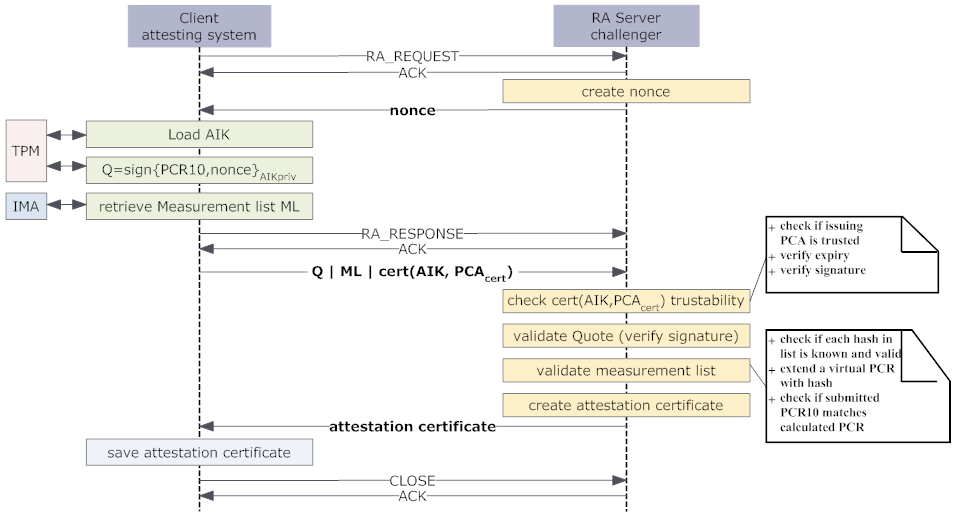}
	\caption{protocol diagram for Remote Attestation}
	\label{fig:small_RAA_protocol}
\end{figure}

The client is separated into different methods representing each state of the protocol. The methods are called subsequently.

\bParams
SRK password & needed to load the AIK\\
AIK password & needed to load the AIK\\
AIK label & the label the to be used AIK is stored under\\
RA server IP address & IP address of the RA server\\
RA server port & port of the RA server, see \nameref{Settings} for default value\\
\eParams

All required parameters can be set from the command-line, a fixed mode is provided via the switch '/f'. The required parameters will then be read from Settings.
		\section{RAserver}
The server is separated into different methods representing each state of the Remote-Attestation protocol. The methods are called subsequently.

$\rightarrow$ see \nameref{RAclient} for a detailed description of the Remote-Attestation protocol
	
	\chapter{jTPMtools}
		\label{c-jTPMtools}
		\textsc{de.fraunhofer.sit.tc.ethemba.jTPMtools}\\\\
		This package contains modified source-code taken from jTPMtools.
		\section{AikUtil}

This class was copied and modified from \textsc{iaik.tc.apps.jtt.aik.AikUtil}.

It is used within \nameref{PCAclient} to create EK-Certificates on-the-fly (which are passed to the
\nameref{PCAserver}). Additionally it is indirectly used within \nameref{PCAserver} and directly
used within \nameref{PrivacyCa} to create AIK-Certificates (which are passed to the \nameref{PCAclient}).

\textbf{Modifications applied:}
\begin{itemize}
	\item changed AIK-Certificate creation to use ethemba's global settings for certificate attributes
	\item changed CA-Certificate creation to use ethemba's global settings for certificate attributes
	\item changed EK-Certificate creation to use ethemba's global settings for certificate attributes
	\item changed PE-Certificate creation to use ethemba's global settings for certificate attributes
\end{itemize}
		\section{Client}

This class was copied and slightly modified from \textsc{iaik.tc.apps.jtt.aik.Client}.

It is used within \nameref{PCAclient} in \textit{state3} to run a \textbf{TPM\_CollateIdentityRequest}
command and in \textit{state4} and \textit{state6} to run a \textbf{TPM\_ActivateIdentity} command.

\textbf{Modifications applied:}
\begin{itemize}
	\item changed visibility of \textit{activateIdentity} from \textcolor{accent3}{\textbf{protected}} to \textcolor{accent2}{\textbf{public}} to have outer access
	\item changed visibility of \textit{collateIdentityReq} from \textcolor{accent3}{\textbf{protected}} to \textcolor{accent2}{\textbf{public}} to have outer access
	\item changed visibility of \textit{overrideEkCertificate} from \textcolor{accent3}{\textbf{protected}} to \textcolor{accent2}{\textbf{public}} to have outer access
\end{itemize}
		\section{PrivacyCa}

This class was copied and modified from \textsc{iaik.tc.apps.jtt.aik.PrivacyCa}.

It is used within \nameref{PCAserver} to process the collateIdentity request blob received from the
\nameref{PCAclient} and wrapping a server-generated nonce inside a TPM blob that can only be accessed
by the TPM itself.

In another state of the AIK-Certification protocol (see \nameref{PCAclient}) the second collateIdentity
request blob received from the PCAclient is being processed and the resulting AIK-Certificate is again
wrapped inside a TPM blob that can only be accessed by the TPM itself.

See \nameref{PCAclient} for detailed information on the AIK-Certification protocol.

\textbf{Modifications applied:}
Two new methods were added to implement a handshake in the AIK-Certification protocol. The state3\_sub method provides the wrapping of the server-generated nonce for the client instead of transmitting the certificate in the first step. In state3\_sub the EK certificate is already validated.

In this design, clients without a valid EK certificate will be rejected at an early stage in the protocol. A later verification of the EK certificate might be advantageous as it decreases the burden on the PCA in case the protocol stops during nonce verification. 
The AIK-Certificate is not issued in this stage. 

The method state5\_sub provides a secure wrapping for the AIK-Certificate. First the AIK-Certificate is generated. A new one-time symmetric \nameref{AES} key is being generated and used to encrypt the certificate. The response consists of three parts:
\begin{enumerate}
\item a symmetric session key and the hash of the AIK public key, both encrypted asymmetrically with the public EK of the client. Via this indirection, only the client is able to decrypt the session key.
\item the symmetric (\nameref{AES}) key previously used to encrypt the AIK-Certificate, encrypted with the session key
\item the AIK-Certificate, encrypted with the \nameref{AES} key
\end{enumerate}

The client can then decrypt the session key using the \textbf{TPM\_ActivateIdentity} function of the TPM. This reveals the \nameref{AES} key, which can then in turn be used to decrypt the AIK-Certificate.
	
	\chapter{modules}
		\textsc{de.fraunhofer.sit.tc.ethemba.modules}\\\\
		This package contains modules providing storage/mapping methods for keys and certificates
		on the one hand and modules providing convenient TPM method handling (e.g. quote, bind, certify etc.) on the other hand.
		\section{client.CertDB}
\subsection{Description}
CertDB saves X509-Certificates in a database mapping it to given UUIDs.

%
% getCert
%
\publicMethod
%title
{getCert}
%description
{returns the corresponding certificate for the given UUID}
%settings
{CertDBfile}
%parameters
{\J{TcTssUuid} uuid & UUID of the certificate}
%output
{\J{X509Certificate} corresponding certificate}

%
% putCert
%
\publicMethodNoOut
%title
{putCert}
%description
{puts a given certificate into the CertDB using the given UUID}
%settings
{CertDBfile}
%parameters
{\J{TcTssUuid} uuid & UUID of the certificate\\
 \J{X509Certificate} cert & certificate to be put into the CertDB}

%
% removeCert
%
\publicMethodNoOut
%title
{removeCert}
%description
{removes a certificate from the CertDB}
%settings
{CertDBfile}
%parameters
{\J{TcTssUuid} uuid & UUID of the certificate}

%
% exportCert
%
\publicMethodNoOut
%title
{exportCert}
%description
{exports a certificate to the disk}
%settings
{CertDBfile}
%parameters
{\J{TcTssUuid} uuid & UUID of the certificate\\
 \jString outputFile & location of the exported certificate}

%
% loadDB
%
\privateMethod
%title
{loadDB}
%description
{loads the CertDB from the globally defined location}
%settings
{CertDBfile}
%parameters
{none}
%output
{\J{Hashtable<String, byte[]>} loaded database}

%
% saveDB
%
\privateMethodNoOut
%title
{saveDB}
%description
{saves a given CertDB to the globally defined location}
%settings
{CertDBfile}
%parameters
{\J{Hashtable<String, byte[]>} db & database to be saved}
		\section{client.CertifyKey}
\label{CK}
\subsection{Description}
This client module can be used to create and certify a TPM key for binding or sealing. The created key is signed by
a given AIK.

\begin{figure}[htbp]
	\centering
		\includegraphics[width=1.00\textwidth]{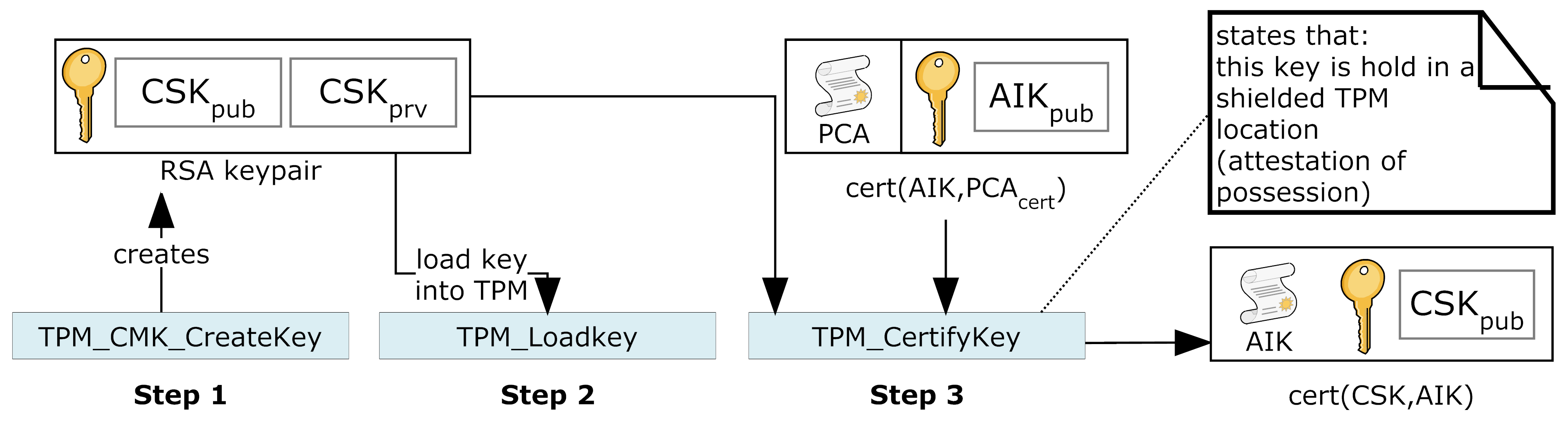}
	\caption{Creation of a CSK (Certified Signing Key) using a certified AIK}
	\label{fig:small_CSKcreation}
\end{figure}

%
% CONSTRUCTOR
%
\constructor
%title
{CertifyKey}
%description
{takes the given input values and stores them internally for later passing it to the TPM via the run method}
%settings
{none}
%parameters
{\J{boolean} isBindingKey & create a binding key (\J{true}) or sealing key (\J{false})\\
 \J{boolean} isVolatile & volatile or non-volatile key\\
 \jString srkPwd & SRK-Password\\
 \jString keyPwd & Key-Password\\
 \jString keyLabel & Label to store created key under\\
 \jInta pcrSelection & array to define, to which PCRs the key shall be bound to (optional)\\
 \J{byte[]} nonce & nonce to be included in certification (anti-replay!)\\
 \jString aikPwd & AIK-Password\\
 \jString aikLabel & Label of AIK to be used for signing}

%
% run
%
\publicMethod
%title
{run}
%description
{Implements the necessary steps to create and certify a TPM key using the TPM. First the SRK (to store the key under) and AIK (to sign the key) are loaded and key attributes for the newly created key are being assigned. Then a composite object is created holding the selected PCRs given via the constructor. This data is passed to the TPM using the certifyKey method provided by jTSS (see \nameref{trustedjava}). After creating/certifying the key, it is stored in the persistent storage and labeled with the given keyLabel. An \J{Object[]} is returned containing
everything needed in \nameref{CKV}:
\begin{itemize}
	\item \J{TcBlobData} public key
	\item \J{TcTssValidation} keyCertification
	\item \J{X509Certificate} AIK-Certificate
\end{itemize}
~
}
%settings
{pwdEncoding, TPM\_KeySize}
%parameters
{none}
%output
{\J{Object[]} containing [1] \J{TcBlobData} public key, [2] \J{TcTssValidation} keyCertification, [3] \J{X509Certificate} AIK-Certificate}

		\section{client.CreateKey}
\subsection{Description}
This client module can be used to create a TPM key for binding or sealing.

%
% CONSTRUCTOR
%
\constructor
%title
{CreateKey}
%description
{takes the given input values and stores them internally for later passing it to the TPM via the run method}
%settings
{none}
%parameters
{\J{boolean} isBindingKey & create a binding key (\J{true}) or sealing key (\J{false})\\
 \J{boolean} isVolatile & volatile or non-volatile key\\
 \J{boolean} isMigratable & migratable or non-migratable key\\
 \jString srkPwd & SRK-Password\\
 \jString keyPwd & Key-Password\\
 \jString keyLabel & Label to store created key under\\
 \jInta pcrSelection & array to define, to which PCRs the key shall be bound to (optional)}

%
% run
%
\publicMethod
%title
{run}
%description
{Implements the necessary steps to create a TPM key using the TPM. First the SRK (to store the key under) is loaded and key attributes for the newly created key are being assigned. Then a composite object is created holding the selected PCRs given via the constructor. This data is passed to the TPM using the createKey method provided by jTSS (see \nameref{trustedjava}). After creating the key, it is stored in the persistent storage and labeled with the given keyLabel.}
%settings
{pwdEncoding, TPM\_KeySize}
%parameters
{none}
%output
{\J{TcBlobData} public part of created keypair}
		\section{client.DataBinding}
\subsection{Description}
This client module can be used to bind (i.e. encrypt) data to the TPM. The bound data can only be unbound (i.e. decrypted) inside this TPM.

%
% CONSTRUCTOR
%
\constructor
%title
{DataBinding}
%description
{takes the given input values and stores them internally for later processing in the run method}
%settings
{none}
%parameters
{\J{byte[]} data & data to be bound/encrypted\\
 \jString srkPwd & SRK-Password\\
 \jString keyPwd & Key-Password\\
 \jString keyLabel & Label of the key to be used to bind the data}

%
% run
%
\publicMethod
%title
{run}
%description
{Implements the necessary steps to bind the given data using the public key belonging to the given keyLabel. First the key is loaded using the SRK. Then data is separated into chunks matching the keysize and passed to the bind method provided by jTSS (see \nameref{trustedjava}).}
%settings
{pwdEncoding, TPM\_KeySize}
%parameters
{none}
%output
{\J{byte[]} bound/encrypted data}
		\section{client.DataUnbinding}
\subsection{Description}
This client module can be used to unbind (i.e. decrypt) once bound data.

%
% CONSTRUCTOR
%
\constructor
%title
{DataUnbinding}
%description
{takes the given input values and stores them internally for later processing in the run method}
%settings
{none}
%parameters
{\J{byte[]} data & data to be bound/encrypted\\
 \jString srkPwd & SRK-Password\\
 \jString keyPwd & Key-Password\\
 \jString keyLabel & Label of the key to be used to unbind the data}

%
% run
%
\publicMethod
%title
{run}
%description
{Implements the necessary steps to unbind the given data using the public key belonging to the given keyLabel. First the key is loaded using the SRK. Then data is separated into chunks matching the keysize and passed to the unbind method provided by jTSS (see \nameref{trustedjava}).}
%settings
{pwdEncoding}
%parameters
{none}
%output
{\J{byte[]} unbound/decrypted data}
		\section{client.QuoteRetrieval}
\subsection{Description}
This client module can be used to request and retrieve a quote from the TPM. The quote includes the signed hash value of the desired PCRs and a signed nonce, as replay protection. The client must provide a valid certificate for the key used to sign the quote. Normally an AIK is used to sign the quote, so a previously acquired AIK certificate from \nameref{PCAclient}/\nameref{PCAserver} can be used to verify the quote.

%
% CONSTRUCTOR
%
\constructor
%title
{QuoteRetrieval}
%description
{takes the given input values and stores them internally for later retrieval of the quote via the run method}
%settings
{none}
%parameters
{\jInta pcrSelection & array to define which PCRs shall be included in the quote\\
 \J{byte[]} nonce & nonce to be included in quote ($\rightarrow$ replay protection)\\
 \jString srkPwd & SRK password, needed to load the AIK\\
 \jString aikPwd & AIK password, needed to load the AIK\\
 \jString aikLabel & label that has been assigned to the AIK in a PCA process\\}

%
% run
%
\publicMethod
%title
{run}
%description
{Implements the necessary steps to retrieve the quote from the TPM. First the AIK is loaded using the SRK. Then a composite
object is created holding the selected PCRs given via the constructor. This data altogether with the nonce is passed to the
TPM using the quote method provided by jTSS (see \nameref{trustedjava}). After retrieving the quote, the AIK is unloaded from
the TPM and the quote is returned to the calling method.}
%settings
{pwdEncoding}
%parameters
{none}
%output
{\J{TcTssValidation} the quote that was retrieved from the TPM}
		\section{client.TpmKeyDB}
\subsection{Description}
TpmKeyDB maps labels to UUIDs. This is used to allow convenient access to keys generated by the TPM.
Those keys are separated by TPM generated UUIDs which are rather difficult to remember. TpmKeyDB allows
to assign labels to those UUIDs and to access them in an easy way.

%
% getUUID
%
\publicMethod
%title
{getUUID}
%description
{returns the corresponding UUID for the given label}
%settings
{TpmKeyDBfile}
%parameters
{\jString label & label to get the UUID for}
%output
{\J{TcTssUuid} corresponding UUID for the given label}

%
% putUUID
%
\publicMethodNoOut
%title
{putUUID}
%description
{puts a label $\Leftrightarrow$ UUID mapping to the TpmKeyDB}
%settings
{TpmKeyDBfile}
%parameters
{\jString label & label to be used\\
 \J{TcTssUuid} uuid & UUID to be used}

%
% removeUUID
%
\publicMethodNoOut
%title
{removeUUID}
%description
{removes a label $\Leftrightarrow$ UUID mapping from the TpmKeyDB}
%settings
{TpmKeyDBfile}
%parameters
{\jString label & label whose mapping is to be removed}

%
% loadDB
%
\privateMethod
%title
{loadDB}
%description
{loads the TpmKeyDB from the globally defined location}
%settings
{TpmKeyDBfile}
%parameters
{none}
%output
{\J{Hashtable<String, String>} loaded database}

%
% saveDB
%
\privateMethodNoOut
%title
{saveDB}
%description
{saves a given TpmKeyDB to the globally defined location}
%settings
{TpmKeyDBfile}
%parameters
{\J{Hashtable<String, String>} db & database to be saved}
		\section{server.CertifyKeyValidaton}
\label{CKV}
\subsection{Description}
This server module can be used to validate a given output of a \nameref{CK} command. The included keyCertification is equipped with a signature and contains a nonce ($\rightarrow$ replay protection). First the signature is checked for validity, then the nonce is matched against the challenged one (see \nameref{CK}). At last the digests of the included public key and the one contained in the output of \nameref{CK} are checked for identicalness.

\textbf{Note:} CertifyKeyValidation only checks the result of a \nameref{CK} command for \textbf{validity}. One should \textbf{verify} the included AIK-Certificate for trustability after successful validation. The included public key may then be used in further steps.

%
% CONSTRUCTOR
%
\constructor
%title
{CertifyKeyValidaton}
%description
{takes the given input values and stores them internally for later validation of the certification via the run method}
%settings
{none}
%parameters
{\J{Object[]} certifyKeyResult & result of the \nameref{CK} command, contains public key, keyCertification and AIK-Certificate\\
 \J{byte[]} nonce & nonce that was challenged and to validate the certification against}

%
% private createDigestInfoDER
%
\privateMethod
%title
{createDigestInfoDER}
%description
{code from jTPM-Tools. It is used to return the digestInfo properly out of the certification.}
%settings
{none}
%parameters
{\textsc{TcBlobData} digest & digest to be properly encoded}
%output
{\textsc{TcBlobData} digestInfo used during verification of the certification signature}

%
% run
%
\publicMethod
%title
{run}
%description
{Implements the necessary steps to validate the given output of a previous \nameref{CK} command.

First the signature of the keyCertification is validated against the AIK-Certificate included in \nameref{CK}'s output. Then the nonce is checked against the given one. Finally the public key included in the output is checked against the one signed in the keyCertification.

\textbf{Note:} Please be aware that this method only validates the output without verifying the trustability of the contained AIK-Certificate. This should be done after a successful validation.}
%settings
{none}
%parameters
{none}
%output
{\J{boolean} \J{true} if certification is valid, else \J{false}}
		\section{server.ExternalDataBinding}
\subsection{Description}
This server module can be used to bind (i.e. encrypt) data to an external TPM using a public key of a TPM keypair resident inside the TPM. Bound data can only be unbound (i.e. decrypted) inside the TPM it was bound to.

%
% CONSTRUCTOR
%
\constructor
%title
{ExternalDataBinding}
%description
{takes the given input values and stores them internally for later processing in the run method}
%settings
{none}
%parameters
{\J{byte[]} data & data to be bound/encrypted\\
 \J{TcBlobData} key & public key to be used to bind data}

%
% run
%
\publicMethod
%title
{run}
%description
{Implements the necessary steps to bind the given data using the given public key. Data is separated into chunks matching the keysize and passed to the bind method provided by jTSS (see \nameref{trustedjava}).}
%settings
{TPM\_KeySize}
%parameters
{none}
%output
{\J{byte[]} bound/encrypted data to be passed to the corresponding TPM}
		\section{server.KeyStorage}

\subsection{Description}
The server module \nameref{server.KeyStorage} provides easy access to save and load functions for server keys. It can be used to store public/private keypairs and provides an interface to retrieve the keys via tags. It is used by \nameref{PCAserver} and \nameref{RAserver} to store their signing keys. If a keypair with the specified tag already exists, the older one will be overwritten.

%
% private loadFromFile
%
\privateMethodNoOut
%title
{loadFromFile}
%description
{loads the mapping file from the predefined filesystem-location. }
%settings
{KeyStorageBaseDir, KeyStorageDB}
%parameters
{none}

%
% private saveToFile
%
\privateMethodNoOut
%title
{saveToFile}
%description
{saves the mapping to the predefined filesystem-location}
%settings
{KeyStorageBaseDir, KeyStorageDB}
%parameters
{none}

%
% private getMapping
%
\privateMethod
%title
{getMapping}
%description
{returns the filename mapped to the given tag}
%settings
{none}
%parameters
{\jString tag & tag to be searched for}
%output
{\jStringa filenames of key files mapped to the tag

[0] public keyfile
[1] private keyfile}

%
% getPublicKeyFile
%
\publicMethod
%title
{getPublicKeyFile}
%description
{returns the filename of the public key}
%settings
{KeyStorageBaseDir}
%parameters
{\jString tag & keypair tag}
%output
{\jString the filename of the public key to the given tag}

%
% getPrivateKeyFile
%
\publicMethod
%title
{getPrivateKeyFile}
%description
{returns the filename of the private key}
%settings
{KeyStorageBaseDir}
%parameters
{\jString tag & keypair tag}
%output
{\jString the filename of the private key to the given tag}

%
% getPublicKey
%
\publicMethod
%title
{getPublicKey}
%description
{get the public key as \textsc{PublicKey} for the given tag}
%settings
{none}
%parameters
{\jString tag & keypair tag}
%output
{\textsc{PublicKey} stored under the given tag}

%
% getPrivateKey
%
\publicMethod
%title
{getPrivateKey}
%description
{get the private key as \textsc{PrivateKey} for the given tag}
%settings
{none}
%parameters
{\jString tag & keypair tag}
%output
{\textsc{PrivateKey} stored under the given tag}

%
% put 
%
\publicMethod
%title
{put}
%description
{put stores the given keypair in the files specified by the user. The user must provide a tag by which the keys can be accessed later. The keys will be stored in Settings.KeyStorageBaseDir, the mapping of tags to keys is stored in Settings.KeyStorageDB.}
%settings
{KeyStorageBaseDir}
%parameters
{\jString tag & the user provided tag for the storage of the keypair\\
\jString publicKeyFile & the filename of the public key file\\
\textsc{PublicKey} publicKey & the public key to be stored\\
\jString privateKeyFile & the filename of the private key file\\
\textsc{PrivateKey} privateKey & the private key to be stored
}
%output
{\jStringa of replaced entries, if tag already existed in \nameref{server.KeyStorage}}

%
% putPublicKey
%
\publicMethod
%title
{putPublicKey}
%description
{stores the given public key in the file specified by the user. The user must provide a tag by which the key can be accessed later. The key will be stored in Settings.KeyStorageBaseDir, the mapping of tags to keys is stored in Settings.KeyStorageDB.}
%settings
{KeyStorageBaseDir}
%parameters
{\jString tag & the user provided tag for the storage of the keypair\\
\jString publicKeyFile & the filename of the public key file\\
\textsc{PublicKey} publicKey & the public key to be stored}
%output
{\jStringa of replaced entries, if tag already existed in \nameref{server.KeyStorage}}

%
% putPublicKey
%
\publicMethod
%title
{putPrivateKey}
%description
{stores the given private key in the file specified by the user. The user must provide a tag by which the key can be accessed later. The key will be stored in Settings.KeyStorageBaseDir, the mapping of tags to keys is stored in Settings.KeyStorageDB.}
%settings
{KeyStorageBaseDir}
%parameters
{\jString tag & the user provided tag for the storage of the keypair\\
\jString privateKeyFile & the filename of the public key file\\
\textsc{PrivateKey} privateKey & the private key to be stored}
%output
{\jStringa of replaced entries, if tag already existed in \nameref{server.KeyStorage}}

		\section{server.QuoteValidation}
\subsection{Description}
The server module \nameref{server.QuoteValidation} can be used to validate a given quote from a TPM. The quote includes the signed hash value of the desired PCRs and a signed nonce ($\rightarrow$ replay protection). The client must provide a valid certificate for the key used to sign the quote. Normally an AIK is used to sign the quote, so a previously acquired AIK certificate from \nameref{PCAclient}/\nameref{PCAserver} can be used to verify the quote.

The validation of the quote consists of two major parts:
\begin{enumerate}
	\item Verification of the signature: the public key used for signing is used to verify the given signature on the nonce and the PCR data.
	\item Verification of quote contents: it is verified that the original nonce, supplied by the server, is included in the client's quote. Furthermore, the quote contains the hash value of all quoted PCRs. A pre-calculated PCR value can be supplied to \nameref{server.QuoteValidation}, so that validation will only succeed if the pre-calculated value matches the quoted PCR value.
\end{enumerate}

%
% CONSTRUCTOR
%
\constructor
%title
{QuoteValidation}
%description
{takes the given input values and stores them internally for later validation of the quote via the run method}
%settings
{none}
%parameters
{
\jInta pcrSelection & array to define which PCRs have been included in the given quote\\
\textsc{TcTssValidation} quote & quote as received from the attesting client\\
\textsc{X509Certificate} aikCert & certificate belonging to the key used for signing. It is provided by the client to retrieve the public key and verify the quote's signature.\\
\textsc{byte[][]} vPCRs & pre-calculated vPCR values are provided as \textsc{byte[]}. Multiple vPCRs can be provided. \\
\jBytea nonce & the anti-replay nonce the server previously provided for the client to be included in the quote
}

%
% private selectPCR
%
\privateMethod
%title
{selectPCR}
%description
{the method is used to return a \textsc{TcBlobData} structure with a \jBytea representation of a selected PCR. The bit at the given index i is set to 1, all other bits are 0. }
%settings
{none}
%parameters
{\textsc{long} index & index of PCR to select}
%output
{\textsc{TcBlobData} containing the \jBytea representation of the selected PCR}

%
% private createDigestInfoDER
%
\privateMethod
%title
{createDigestInfoDER}
%description
{code from jTPM-Tools. It is used to return the digestInfo properly out of the quote.}
%settings
{none}
%parameters
{\textsc{TcBlobData} digest & digest to be properly encoded}
%output
{\textsc{TcBlobData} digestInfo used during verification of the quote signature}

%
% run
%
\publicMethod
%title
{run}
%description
{Implements the necessary steps to validate the given quote. First the public key is extracted from the supplied certificate. It is used to verify the quote's signature. A composite object with the given pre-calculated PCR values is created and its hash value is calculated. The hash is checked against the hash value provided by the quote. Finally the quote's nonce is checked. Only if all steps of the validation succeed, the quote will validate.
Protection is provided against different attacks. If quote contents are changed after signing, the validation of the signature will fail. If a malicious client provides a modified measurement list, pretending to run only trusted software, the calculated vPCR will not match the quoted and signed PCR value. Anti-replay attack protection is given by the usage of the rolling nonce.}
%settings
{none}
%parameters
{none}
%output
{\J{boolean} \J{true} if quote is valid, else \J{false}}

	\chapter{net}
		\textsc{de.fraunhofer.sit.tc.ethemba.net}\\\\
		This package contains networking classes for sending and receiving data via TCP/IP.
		\section{NetEntity}

\subsection{Description}
NetEntity provides an implementation of a networked client-server infrastructure. NetEntity can be initialized as server or as client. Once initialized, it can be used to transfer controls (\nameref{NetCommand}) or Objects.

%
% Netentity
%
\constructor
%title
{NetEntity}
%description
{Sets up a NetEntity as either server or client. If invoking with serverHostname \textbf{and} serverPort, a client version is instantiated. If \textbf{only} serverPort is given, a server is instantiated}
%settings
{none}
%parameters
{\jString serverHostname & hostname of server the client will connect to (client only)\\
 \jInt serverPort & port the client will connect to / the server will listen on}

%
% init 
%
\publicMethodNoOut
%title
{init}
%description
{Initializes the connection. A socket is created and bound to the specified port. If instance is a server, the socket will start to listen and accept connections.}
%settings
{none}
%parameters
{none}

%
% close 
%
\publicMethodNoOut
%title
{close}
%description
{closes the socket. Returns \J{true}, if socket closed successfully, else \J{false}.}
%settings
{none}
%parameters
{none}

%
% sendACK 
%
\publicMethod
%title
{sendACK}
%description
{sends a \nameref{NetCommand}.ACK}
%settings
{none}
%parameters
{none}
%output
{\J{true}, if sent successfully, else \J{false}}

%
% sendNACK 
%
\publicMethod
%title
{sendNACK}
%description
{sends a \nameref{NetCommand}.NACK}
%settings
{none}
%parameters
{none}
%output
{\J{true}, if sent successfully, else \J{false}}

%
% sendCommand
%
\publicMethod
%title
{sendCommand}
%description
{sends the given \nameref{NetCommand}}
%settings
{none}
%parameters
{\nameref{NetCommand} netCommand & the \textsc{NetCommand} to be sent}
%output
{\J{true}, if sent successfully, else \J{false}}

%
% sendObject 
%
\publicMethod
%title
{sendObject}
%description
{sends the given object over the connection.}
%settings
{none}
%parameters
{\textsc{Object} o & object to send}
%output
{\J{true} if sent successfully, else \J{false}}

%
% receiveACK 
%
\publicMethod
%title
{receiveACK }
%description
{try to receive a \nameref{NetCommand}.ACK}
%settings
{none}
%parameters
{none}
%output
{\J{true} if ACK received, else \J{false}}

%
% receiveCommand
%
\publicMethod
%title
{receiveCommand}
%description
{receive a \nameref{NetCommand}}
%settings
{none}
%parameters
{none}
%output
{returns the received \nameref{NetCommand}, if received object is an integer. Else \nameref{NetCommand}.UNKNOWN is returned}

%
% receiveObject
%
\publicMethod
%title
{receiveObject}
%description
{waits to receive an Object of the given class.}
%settings
{none}
%parameters
{\textsc{Class} c & class of the object to be received}
%output
{If received object is of specified type, the object is returned, else \textsc{null}.}

%
% getRemoteIP
%
\publicMethod
%title
{getRemoteIP}
%description
{returns the IP address of the client, when issued on the server}
%settings
{none}
%parameters
{none}
%output
{\textsc{String} IP address of client}

%
% getRemoteHostname
%
\publicMethod
%title
{getRemoteHostname}
%description
{returns the hostname of the client, when issued on the server}
%settings
{none}
%parameters
{none}
%output
{\textsc{String} hostname of client}

		\section{NetCommand}

\subsection{Description}
NetCommand provides a mapping between symbolic names and command codes (\textsc{int}) for convenient use of \nameref{NetEntity} objects.

\begin{itemize}
	\item \J{UNKNOWN}
	\item \J{ACK}
	\item \J{NACK}
	\item \J{INIT}
	\item \J{CLOSE}
	\item \J{KNOWNHASHES}
	\item \J{STRING}
	\item \J{RA\_REQUEST}
	\item \J{RA\_RESPONSE}
	\item \J{PCA\_REQUEST}
	\item \J{PCA\_RESPONSE}
\end{itemize}

	\chapter{types}
		\textsc{de.fraunhofer.sit.tc.ethemba.types}\\\\
		This package contains data types for convenient access to IMA measurements and hash lists.
		\section{MeasurementList}
\subsection{Description}
MeasurementList is a data type for convenient conversion and access to IMA measurement data.

%
% CONSTRUCTOR
%
\constructor
%title
{MeasurementList}
%description
{initializes the MeasurementList with a \J{String[][][]}}
%settings
{none}
%parameters
{\J{String[][][]} measurementList & [1] PCR number

[2] hash

[3] application}

%
% getMeasurementList
%
\publicMethod
%title
{getMeasurementList}
%description
{returns the MeasurementList as string array representation}
%settings
{none}
%parameters
{none}
%output
{\J{String[][][]} string array representation of the MeasurementList}

%
% getMeasurementListForPCR
%
\publicMethod
%title
{getMeasurementListForPCR}
%description
{returns the MeasurementList for a given PCR number as string array representation}
%settings
{none}
%parameters
{\jInt pcr & to be filtered PCR number}
%output
{\J{String[][]} string array representation of the MeasurementList for the given PCR number}

%
% loadFromFile
%
\publicMethod
%title
{loadFromFile}
%description
{loads MeasurementList from a file in IMA-Measurement-Format}
%settings
{none}
%parameters
{\jString filename & file to load MeasurementList from}
%output
{\J{MeasurementList} loaded MeasurementList}
		\section{KnownHashesList}
\label{KHL}
\subsection{Description}
KnownHashesList is a data type providing basic database abilities for storing known
hash $\Leftrightarrow$ application mappings.

%
% CONSTRUCTOR
%
\constructor
%title
{KnownHashesList}
%description
{initializes the KnownHashesList}
%settings
{none}
%parameters
{none}

%
% put
%
\publicMethod
%title
{put}
%description
{maps the given hash to the given tag}
%settings
{none}
%parameters
{\jString sha1hash & hash of a application\\
 \jString tag & path and name of the application that was hashed}
%output
{\jString tag/application that was associated to the given tag before, else \J{null}}

%
% get
%
\publicMethod
%title
{get}
%description
{returns the tag to which the specified hash is mapped to}
%settings
{none}
%parameters
{\jString sha1hash & hash to be searched for}
%output
{\jString tag/application that hash is associated to, else \J{null}}

%
% containsTag
%
\publicMethod
%title
{containsTag}
%description
{tests if the given tag/application is mapped to one ore more hashes}
%settings
{none}
%parameters
{\jString tag & tag/application to be searched for}
%output
{\J{true} if given tag/application is associated with one or hashes, else \J{false}}

%
% containsSha1Hash
%
\publicMethod
%title
{containsSha1Hash}
%description
{tests if the given hash is mapped to a tag/application}
%settings
{none}
%parameters
{\jString sha1hash & hash to be searched for}
%output
{\J{true} if given hash is associated with a tag/application, else \J{false}}

%
% contains
%
\publicMethod
%title
{contains}
%description
{tests if a given hash is mapped to a given tag/application}
%settings
{none}
%parameters
{\jString sha1hash & hash to be searched for\\
 \jString tag & tag/application to be associated with}
%output
{\J{true} if given hash is associated with the given tag/application, else \J{false}}

%
% saveToFile
%
\publicMethodNoOut
%title
{saveToFile}
%description
{saves KnownHashesList to a given file}
%settings
{none}
%parameters
{\jString filename & file to save KnownHashesList in}

%
% loadFromFile
%
\publicMethod
%title
{loadFromFile}
%description
{loads a KnownHashesList from the disk}
%settings
{none}
%parameters
{\jString filename & file to load KnownHashesList from}
%output
{\J{KnownHashesList} loaded KnownHashesList or \J{null} if KnownHashesList could not be loaded}
	
	\chapter{utils}
		\textsc{de.fraunhofer.sit.tc.ethemba.utils}\\\\
		This package contains classes and methods for data conversion and standardized algorithms,
		that may be accessed in a convenient way.
		\section{AES}
\subsection{Description}
AES provides access to various AES calculations (encryption/decryption). Each method may also be accessed
in a static way. Be aware to pass enough data to the static methods (i.e. AES-Key and IV).

Additionally random and pseudo-random AES-Keys and IVs may be generated using this class.

%
% CONSTRUCTOR
%
\constructor
%title
{AES}
%description
{constructs an AES object with the given key and initialization vector (IV)}
%settings
{none}
%parameters
{\J{SecretKey} k & AES key\\
 \J{byte[]} IV & initialization vector (IV)}

%
% encryptObject
%
\publicMethod
%title
{encryptObject}
%description
{encrypts a given \jObject}
%settings
{none}
%parameters
{\jObject m & object to be encrypted}
%output
{\J{byte[]} encrypted data}

%
% encrypt
%
\publicMethod
%title
{encrypt}
%description
{encrypts a given \J{byte[]}}
%settings
{none}
%parameters
{\J{byte[]} m & data to be encrypted}
%output
{\J{byte[]} encrypted data}

%
% decryptObject
%
\publicMethod
%title
{decryptObject}
%description
{decrypts a given \J{byte[]} back into an \jObject}
%settings
{none}
%parameters
{\J{byte[]} c & data to be decrypted}
%output
{\jObject decrypted object}

%
% decrypt
%
\publicMethod
%title
{decrypt}
%description
{decrypts a given \J{byte[]}}
%settings
{none}
%parameters
{\J{byte[]} c & data to be decrypted}
%output
{\J{byte[]} decrypted data}

%
% crypt
%
\privateMethod
%title
{crypt}
%description
{en- or decrypts given \J{byte[]} with given AES-Key and IV}
%settings
{aesCipherMode, aesKeySize}
%parameters
{\jInt optMode & operation mode: \textit{Cipher.DECRYPT\_MODE} or \textit{Cipher.ENCRYPT\_MODE}\\
 \J{SecretKey} k & AES-Key to be used\\
 \J{byte[]} IV & IV to be used\\
 \J{byte[]} input & data to be en- or decrypted}
%output
{\J{byte[]} en- or decrypted data}

%
% generateKey
%
\publicMethod
%title
{generateKey}
%description
{generates a random AES-Key}
%settings
{none}
%parameters
{none}
%output
{\J{SecretKey} generated key}

%
% generateKey
%
\publicMethod
%title
{generateKey}
%description
{generates a pseudo-random AES-Key by using an initial seed}
%settings
{none}
%parameters
{\J{byte[]} seed & initial seed for PRNG}
%output
{\J{SecretKey} generated key}

%
% generateIV
%
\publicMethod
%title
{generateIV}
%description
{generates a pseudo-random IV by using an initial seed}
%settings
{none}
%parameters
{\J{byte[]} seed & initial seed for PRNG}
%output
{\J{byte[]} generated IV}
		\section{Helpers}
\subsection{Description}
Helpers provides some useful methods for data conversion and serialization.

%
% byteToHexString
%
\publicMethod
%title
{byteToHexString}
%description
{converts a given \J{byte[]} into a Hex-String}
%settings
{none}
%parameters
{\J{byte[]} data & data to be converted}
%output
{\jString Hex-String representation of given data}

%
% hexStringToByteArray
%
\publicMethod
%title
{hexStringToByteArray}
%description
{converts a given Hex-String into a \J{byte[]}}
%settings
{none}
%parameters
{\jString hexstring & string to be converted}
%output
{\J{byte[]} converted data}

%
% leadingZeroes
%
\publicMethod
%title
{leadingZeroes}
%description
{adds leading zeroes to a given number}
%settings
{none}
%parameters
{\jInt num & number to be padded with zeroes\\
 \jInt length & length of result}
%output
{\jString string representation of given number with padded zeroes to match the given length}

%
% saveObjectToFile
%
\publicMethodNoOut
%title
{saveObjectToFile}
%description
{saves a given object (needs to be serializable) to the disk}
%settings
{none}
%parameters
{\jObject o & object to be saved\\
 \jString filename & location of saved object}
 
%
% saveBytesToFile
%
\publicMethodNoOut
%title
{saveBytesToFile}
%description
{saves bytes to a file}
%settings
{none}
%parameters
{\J{byte[]} b & bytes to be saved\\
 \jString filename & file to be saved to}

%
% loadObjectFromFile
%
\publicMethod
%title
{loadObjectFromFile}
%description
{loads an object from a given file}
%settings
{none}
%parameters
{\jString filename & file to load object from}
%output
{\jObject loaded from given file}

%
% loadBytesFromFile
%
\publicMethod
%title
{loadBytesFromFile}
%description
{loads a file byte-wise}
%settings
{none}
%parameters
{\jString filename & file to be loaded from}
%output
{\J{byte[]} loaded bytes}

%
% Object2Bytes
%
\publicMethod
%title
{Object2Bytes}
%description
{converts an object to bytes}
%settings
{none}
%parameters
{\jObject o & object to be converted}
%output
{\J{byte[]} converted bytes}

%
% Bytes2Object
%
\publicMethod
%title
{Bytes2Object}
%description
{converts a bytes back to an object}
%settings
{none}
%parameters
{\J{byte[]} b & bytes to be converted}
%output
{\jObject converted object}
		\section{SHA1}
\subsection{Description}
SHA1 provides access to various SHA1 calculations.

%
% main
%
\publicMethod
%title
{main}
%description
{generates SHA1-Hash of given string/hex-string}
%settings
{none}
%parameters
{\jStringa args & [0] String to be hashed

[1] if '-h' the given String is assumed to be a hex representation}
%output
{SHA1-Hash of given data}

%
% hashByteToByte
%
\publicMethod
%title
{hashByteToByte}
%description
{calculates SHA1-Hash of given \textsc{byte[]}}
%settings
{none}
%parameters
{\J{byte[]} input & data to be hashed}
%output
{\J{byte[]} SHA1-Hash of given data}

%
% hashHex
%
\publicMethod
%title
{hashHex}
%description
{calculate SHA1-Hash for given Hex-String}
%settings
{none}
%parameters
{\jString input & data to be hashed (Hex-String)}
%output
{\jString SHA1-Hash of given data}

%
% hash
%
\publicMethod
%title
{hash}
%description
{calculate SHA1-Hash for given String}
%settings
{none}
%parameters
{\jString input & data to be hashed}
%output
{\jString SHA1-Hash of given data}

%
% randomHash
%
\publicMethod
%title
{randomHash}
%description
{returns SHA1-Hash of a random generated number}
%settings
{none}
%parameters
{none}
%output
{\J{byte[]} of SHA1-Hash}

	\chapter{Demonstrators}
		For demonstration purposes we provide two simple yet useful demonstrators. These demonstrators will
prove that the \textbf{Remote Attestation Protocol} developed and implemented in \textbf{ethemba}
serves as reliable attestation method.

To prove reliability the following demonstrators inject bad code into an application that was added
to the \nameref{RAserver}'s \nameref{KHL}. This KnownHashesList acts as a white-list representing
SHA1 hashes for well-known and valid applications. If an application was run on the attesting system
that either is not included in the KnownHashesList or whose SHA1 hash does not match the one inside
the KnownHashesList, Remote Attestation fails.

To prove this we provide a demonstrator that will compile a C-program out of two sources. The demonstrators
can be found in the subfolder \textbf{demo} and contain the following files:

\begin{itemize}
	\item demoevil.sh
	\item demogood.sh
	\item helloworldevil.c
	\item helloworldgood.c
\end{itemize}

\section{demogood.sh}
This demonstrator will compile the code contained in \textbf{helloworldgood.c} to the binary
\textbf{helloworld}.

\begin{lstlisting}[caption=helloworldgood.c]
#include <stdio.h>

int main()
{
	printf("sawubona!\n");
	printf("means 'hello' in zulu\n");
	return 0;
}
\end{lstlisting}

We assume this code to be \enquote{good} and therefore included its SHA hash into the RAserver's
KnownHashesList. Executing this program will not break a successful Remote Attestion.

\section{demoevil.sh}
This demonstrator will compile the code contained in \textbf{helloworldevil.c} to the binary
\textbf{helloworld}.

\begin{lstlisting}[caption=helloworldevil.c]
#include <stdio.h>

int main()
{
	printf("hamba kahle!\n");
	printf("means 'goodbye' in zulu\n");
	return 0;
}
\end{lstlisting}

This can be seen as an attack pretending execution of a well-known program. As its SHA1 hash now
differs from the one included in the RAserver's KnownHashesList, Remote Attestion now \textbf{has to fail}.
	
	\chapter{Settings}
		\label{Settings}
		%
% Preamble
%
\renewcommand{\nompreamble}{%
Global settings, parameters and constants of ethemba framework. These variables may be
used to customize ethemba's behaviour to individual needs.\\
\\
\textbf{Note:} some variables have inner dependencies.
\\
\textbf{Note:} filesystem related variables must fit OS's naming conventions.

}

%
% Settings
%

\nomenclature{ethembaDir}
{root directory for ethemba}
\nomenclature{IMAruntimeFile}
{location of IMA-Measurement-Log in filesystem}

\nomenclature{PCAServerIP}
{PCA-Server's IP-Address (used in fixed-mode of \nameref{PCAclient})}
\nomenclature{PCAServerPort}
{PCA-Server's Port}
\nomenclature{PCAServer\_KeyTag}
{default label for Server-Key-Storage}
\nomenclature{PCAdefault\_AIKtag}
{default label for AIK-Creation/-Storage (used in fixed-mode of \nameref{PCAclient})}
\nomenclature{PCAcertCountry}
{PCA-Certificate Attribute: Country}
\nomenclature{PCAcertOrganization}
{PCA-Certificate Attribute: Organization}
\nomenclature{PCAcertOU}
{PCA-Certificate Attribute: OU}
\nomenclature{PCAcertCommonName}
{PCA-Certificate Attribute: CommonName}
\nomenclature{PCAcertSerialNumber}
{PCA-Certificate Attribute: SerialNumber}
\nomenclature{PCAcertPrivateKeySize}
{PCA-Certificate Attribute: PrivateKeySize}
\nomenclature{PCAcertPolicyOID}
{PCA-Certificate Attribute: PolidyOID}
\nomenclature{PCAcertPolicyURL}
{PCA-Certificate Attribute: PolicyURL}

\nomenclature{RAServerIP}
{RA-Server's IP-Address (used in fixed-mode of \nameref{RAclient})}
\nomenclature{RAServerPort}
{RA-Server's Port}
\nomenclature{RAServer\_KeyTag}
{default label for Server-Key-Storage}
\nomenclature{RAdefault\_AIKtag}
{default label for AIK-Creation/-Storage (used in fixed-mode of \nameref{RAclient})}
\nomenclature{RAServer\_KnownHashesList}
{filename of stored KnownHashesList (used by \nameref{RAserver}).}
\nomenclature{RAcert\_Expiry}
{expiry of Remote Attestation Certificates (relative to server-time)}
\nomenclature{RAcertCountry}
{RA-Certificate Attribute: Country}
\nomenclature{RAcertOrganization}
{RA-Certificate Attribute: Organization}
\nomenclature{RAcertOU}
{RA-Certificate Attribute: OU}
\nomenclature{RAcertCommonName}
{RA-Certificate Attribute: CommonName}

\nomenclature{OwnerPwd}
{default TPM-Owner password (used in fixed-mode of \nameref{PCAclient} and \nameref{RAclient})}
\nomenclature{SRKPwd}
{default SRK password (used in fixed-mode of \nameref{PCAclient} and \nameref{RAclient})}
\nomenclature{AIKPwd}
{default AIK password (used in fixed-mode of \nameref{PCAclient} and \nameref{RAclient})}

\nomenclature{pwdEncoding}
{password encoding used for TPM-Key passwords}
\nomenclature{TPM\_CLIcertCountry}
{TPM-Client-Certificate Attribute: Country}
\nomenclature{TPM\_CLIcertOrganization}
{TPM-Client-Certificate Attribute: Organization}
\nomenclature{TPM\_CLIcertOU}
{TPM-Client-Certificate Attribute: OU}
\nomenclature{TPM\_CLI\_PE\_policyOID}
{PE-Certificate Attribute: Policy OID}
\nomenclature{TPM\_CLI\_PE\_policyURL}
{PE-Certificate Attribute: Policy URL}
\nomenclature{TPM\_CLI\_PE\_Pmanufacturer}
{PE-Certificate Attribute: Platform Manufacturer}
\nomenclature{TPM\_CLI\_PE\_Pmodel}
{PE-Certificate Attribute: Platform Model}
\nomenclature{TPM\_CLI\_PE\_Pversion}
{PE-Certificate Attribute: Platform Version}
\nomenclature{TPM\_CLI\_PE\_Pclass}
{PE-Certificate Attribute: Platform Class}
\nomenclature{TPM\_CLI\_PE\_majorVersion}
{PE-Certificate Attribute: Platform Major Version}
\nomenclature{TPM\_CLI\_PE\_minorVersion}
{PE-Certificate Attribute: Platform Minor Version}
\nomenclature{TPM\_CLI\_PE\_revision}
{PE-Certificate Attribute: Platform Revision}
\nomenclature{TPM\_CLI\_EK\_policyOID}
{EK-Certificate Attribute: Policy OID}
\nomenclature{TPM\_CLI\_EK\_policyURL}
{EK-Certificate Attribute: Policy URL}
\nomenclature{TPM\_CLI\_EK\_TPMmanufacturer}
{EK-Certificate Attribute: TPM Manufacturer}
\nomenclature{TPM\_CLI\_EK\_TPMmodel}
{EK-Certificate Attribute: TPM Model}
\nomenclature{TPM\_CLI\_EK\_TPMversion}
{EK-Certificate Attribute: TPM Version}
\nomenclature{TPM\_CLI\_EK\_SpecFamily}
{EK-Certificate Attribute: TPM-Spec-Family}
\nomenclature{TPM\_CLI\_EK\_SpecLevel}
{EK-Certificate Attribute: TPM-Spec-Level}
\nomenclature{TPM\_CLI\_EK\_SpecRevision}
{EK-Certificate Attribute: TPM-Spec-Revision}
\nomenclature{TPM\_CLI\_AIK\_policyOID}
{AIK-Certificate Attribute: Policy OID}
\nomenclature{TPM\_CLI\_AIK\_policyURL}
{AIK-Certificate Attribute: Policy URL}

\nomenclature{TpmKeyDBfile}
{location of TPM-Key-DB}
\nomenclature{KeyStorageBaseDir}
{location of Server-Key-Storage}
\nomenclature{KeyStorageDB}
{location of Server-Key-Storage-DB}
\nomenclature{ServerKeyAlgorithm}
{Server-KeyPair's Algorithm (used in \nameref{CreateServerKeyPair})}
\nomenclature{ServerKeySize}
{Server-KeyPair's Key-Size}
\nomenclature{ServerSignAlgorithmID}
{Algorithm used for signing}

\nomenclature{CertDBfile}
{location of Certificate-DB}
\nomenclature{CertExportBaseDir}
{location of exported certificates}

\nomenclature{aesKeySize}
{key-size for AES-Keys (used in \nameref{AES})}
\nomenclature{aesCipherMode}
{AES-Cipher-Mode (used in \nameref{AES})}

%\nomenclature{}
%{}

\markboth{\nomname}{\nomname}
\printnomenclature		

	\chapter*{References}
		\addcontentsline{toc}{chapter}{References}
		\ethemba uses \textbf{jTSS} as underlying implementation of the TCG Software Stack
for Java.

jTSS implements all the TSS layers directly in Java by staying very close to the TPM
specifications stated by the TCG. It is developed and maintained at the Institute for
Applied Information Processing and Communication (Institut für angewandte
Informationsverarbeitung und Kommunikation, IAIK) at Graz University of Technology (TU Graz).

IAIK additionally provides a set of command line tools for basic operations on TPMs. These
tools include several modules for attesting a platform (e.g. generate AIKs, certificates and handle TPM quotes).

Some code of \textbf{jTPMtools} was re-used in \ethemba to allow for convenient and reliable
access. These modified modules can be found in a separate package underneath ethemba (\nameref{c-jTPMtools}).

~\\

For further information on jTSS and the OpenTC Project visit the following websites:

\textbf{Trusted Computing for the Java Platform}\\
\label{trustedjava}
\url{http://trustedjava.sourceforge.net}

\textbf{OpenTC Project}\\
\label{OpenTC}
\url{http://www.opentc.net}

\textbf{Institute for Applied Information Processing and Communications}\\
\label{IAIK}
\url{http://www.iaik.tugraz.at}

	\clearpage
	\thispagestyle{empty}
\setlength{\evensidemargin}{0.43cm}

~

\vspace{30em}

\begin{center}
	\large \ethembaG
\end{center}

\vfill
\begin{center}
	\textit{\LARGE ...just in case you didn't know...}
\end{center}

\end{document}